\documentclass{eas}
\usepackage{graphicx}
\newcommand\degr{\hbox{$^\circ$}}

\begin{document}

\title{Forbidden calcium lines as disc tracers} 
\author{Anna Aret}\address{Tartu Observatory, 61602, T\~oravere, Tartumaa, Estonia}
\author{Michaela Kraus}\sameaddress{1} 
\secondaddress{Astronomick\'y \'ustav AV \v{C}R,
Fri\v{c}ova 298, 251\,65 Ond\v{r}ejov, Czech Republic}
\begin{abstract}
Forbidden emission lines are particularly valuable disc tracers, because their profiles reflect the kinematics within their formation region.
Here we present a short excerpt from the results of a spectro\-scopic survey of evolved massive stars surrounded by high-density discs.
\end{abstract}
\maketitle
\section{Introduction}

Emission line stars are typically surrounded by large amount of dense 
circumstellar material (CSM) that often accumulates in rings or disc-like structures. 
Gas diagnostics using forbidden [O\,{\sc i}] lines is well known: these lines arise in high-density environments, such as the 
inner disc regions around B[e] supergiants. Recently we have discovered also [Ca\,{\sc ii}] $\lambda\lambda$7291, 7324 lines in the spectra of B[e] supergiants, which trace  hotter regions closer to the star than the [O\,{\sc i}] lines (\cite{2012MNRAS.423..284A}).
We initiated a survey of emission
line stars in different evolutionary phases with various 
circumstellar environments to study physical conditions in the CSM.
\begin{table}[h]
\caption{Stellar parameters and extracted disc kinematics}
\begin{tabular}{lclc|llll}
\hline
               &                 &           &      &\multicolumn{2}{c}{[Ca\,{\sc ii}]}       &\multicolumn{2}{c}{[O\,{\sc i}]}         \\
Star           &  $M_{*}$        &  $i^{a}$  &Ref   & ~~$v_{\rm rot}$ &~~~~$R$                & ~~$v_{\rm rot}$  &~~~~$R$              \\
               &  [$M_{\odot}$]  &  [\degr]  &      & [km/s]          &~~[AU]                 & [km/s] &~~[AU]                          \\
\hline                    
V1478 Cyg      &  38--40         & 82        &(1)   & 38$\pm$1        &24.6$\pm$1.3           & 25$\pm$1         &55.7$\pm$5.9           \\
l Pup          &  15--20         & 38        &(2)   & 72$\pm$1        &\phantom{0}3.0$\pm$0.4 & 68$\pm$1         &\phantom{0}3.4$\pm$0.5 \\
V1429 Aql      & 66$\pm$9        & 73$\pm$13 &(3)   & 50$\pm$1        &23.4$\pm$4.0           &  \phantom{00}--  &\phantom{00.0}--       \\
OY Gem         & 0.62            & --        &(4)   & \phantom{0}0    &\phantom{00.0}--       & \phantom{0}0     &\phantom{00.0}--        \\
\hline
\end{tabular}

\small
$^{a}$Disc inclination $i$ of 90\degr means edge-on view of the disc. \\
(1)~B{\' a}ez-Rubio A., {\etal} 2013, A\&A, 553, A45;
(2)~Millour F., {\etal} 2011, A\&A, 526, A107;
(3)~Lobel A., {\etal} 2013, A\&A, 559, A16;
(4)~Arkhipova V. P., Ikonnikova N. P., 1992, Soviet Ast. Lett., 18, 418
\end{table}

\section{Observations and modelling}
The observations of the [O\,{\sc i}] $\lambda$6300 and the [Ca\,{\sc ii}] $\lambda\lambda$7291, 7324 lines
were obtained using the Coud{\'e} spectrograph attached to the \mbox{2-m} telescope at Ond{\v r}ejov Observatory (\cite{2002PAICz..90....1S}) with a spectral resolution of $\sim 20$\,km\,s$^{-1}$.

We applied a simple, purely kinematic model. Assuming the emission originates 
from a narrow Keplerian rotating ring,
we calculated the profile shape considering only the rotational velocity,
projected to the line of sight according to the observed inclination
angles (Table~1), and the resolution of the spectrograph.

\section{Results}
The good fits to the observed line profiles (Figure~1) demonstrate that the observed emission originates indeed from a narrow ring region
with radius $R$ (Table~1).
For the two B[e] supergiants, V1478 Cyg and l Pup, the kinematics
obtained from the [O\,{\sc i}] and [Ca\,{\sc ii}] line profiles agrees with an origin of the lines in the Keplerian rotating disc. 
The LBV candidate V1429 Aql shows no [O\,{\sc i}] lines, but the profile of its [Ca\,{\sc ii}] 
lines suggests that the emission originates in its hot, ionized circumbinary disc.
The forbidden lines in the spectra of the compact planetary nebula OY Gem display no kinematical broadening beyond spectral resolution. 

\begin{figure}
\centering
\includegraphics[width=0.8\textwidth]{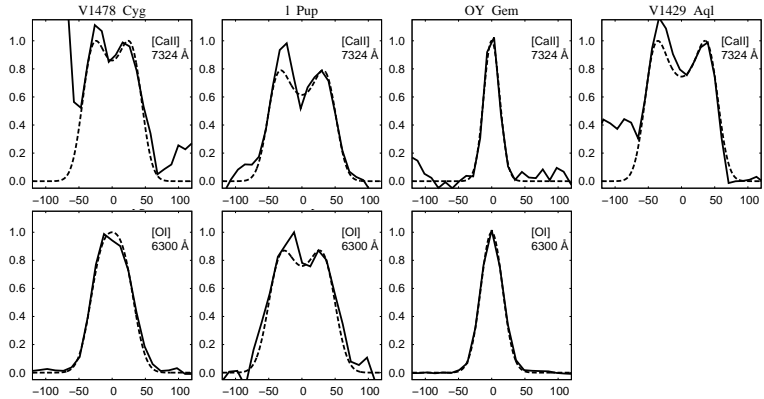}
\caption{Model fits (dashed) to the observed profiles (solid) of the forbidden lines.}
\end{figure}

\medskip
\textbf{Acknowledgements:} 
A.A. acknowledges financial support from Estonian grants ETF8906 and IUT40-1; 
M.K. from GA\,\v{C}R (14-21373S) and  RVO:67985815. 


\end{document}